\documentclass[11pt,a4paper]{article}
\usepackage[utf8]{inputenc}
\usepackage{jheppub}
\usepackage{graphicx}
\usepackage{dcolumn}
\usepackage{bm}
\usepackage{color}
\usepackage[dvipsnames]{xcolor}
\usepackage{url}
\usepackage{float}
\usepackage{amsmath,amssymb,graphicx}
\usepackage{amsmath}
\usepackage{multirow}
\usepackage{amssymb}
\usepackage{mathrsfs}
\usepackage{comment}
\usepackage{accents}
\usepackage{epsfig}
\usepackage[lofdepth,lotdepth,caption=false]{subfig}
\usepackage{mathtools}
\usepackage{mathrsfs}
\usepackage{mathtools}
\usepackage{booktabs}
\usepackage{slashbox}
\usepackage{adjustbox}

\DeclarePairedDelimiterX\braket[2]{\langle}{\rangle}{#1 \delimsize\vert #2}

\newcommand{\beq}{\begin{equation}}
\newcommand{\eeq}{\end{equation}}
\newcommand{\beqa}{\begin{eqnarray}}
\newcommand{\eeqa}{\end{eqnarray}}
\newlength\figureheight 
\newlength\figurewidth 
\definecolor{MyGrey}{rgb}{0,0,0} 
\definecolor{MyDarkBlue}{rgb}{0.3,0.3,0.9} 
\definecolor{MyLightBlue}{rgb}{0.22,0.51,0.9}
\usepackage{cancel}

\def\beq{\begin{equation}}
\def\eeq{\end{equation}}

\title{A Short Travel for Neutrinos in Large Extra Dimensions}
\author{G.  V. Stenico$^{1,2}$,}

\author{D. V. Forero$^{1}$}

\author{and O. L. G. Peres$^{1}$}

\affiliation{
$^{1}$Instituto de F\'isica Gleb Wataghin - UNICAMP, 13083-859, Campinas, SP, Brazil \\
$^{2}$Northwestern University, Department of Physics \& Astronomy, 2145 Sheridan Road, Evanston, IL 60208, USA}
\emailAdd{gstenico@ifi.unicamp.br}
\emailAdd{dvanegas@ifi.unicamp.br}
\emailAdd{orlando@ifi.unicamp.br}
\date{\today} 

\abstract{Neutrino oscillations successfully explain the flavor transitions observed in neutrinos produced in natural sources like the center of the sun and the earth atmosphere, and also from man-made sources like reactors and accelerators. These oscillations are driven by two mass-squared differences, solar and atmospheric, at the sub-eV scale. However, longstanding anomalies at short-baselines might imply the existence of new oscillation frequencies at the eV-scale and the possibility of this sterile state(s) to mix with the three active neutrinos. One of the many future neutrino programs that are expected to provide a final word on this issue is the Short-Baseline Neutrino Program (SBN) at FERMILAB. In this letter, we consider a specific model of Large Extra Dimensions (LED) which provides interesting signatures of oscillation of extra sterile states. We started re-creating sensitivity analyses for sterile neutrinos in the 3+1 scenario, previously done by the SBN collaboration, by simulating neutrino events in the three SBN detectors from both muon neutrino disappearance and electron neutrino appearance. Then, we implemented neutrino oscillations as predicted in the LED model and also we have performed sensitivity analysis to the LED parameters. Finally, we studied the SBN power of discriminating between the two models, the 3+1 and the LED. We have found that SBN is sensitive to the oscillations predicted in the LED model and have the potential to constrain the LED parameter space better than any other oscillation experiment, for $m_{1}^D<0.1\,\text{eV}$. In case SBN observes a departure from the three active neutrino framework, it also has the power of discriminating between sterile oscillations predicted in the 3+1 framework and the LED ones. }
\keywords{neutrino oscillation, large extra dimension, short-baseline}
\begin{document}
\maketitle

\section{Introduction} 

Our knowledge of the three neutrino oscillation paradigm has substantially improved in the last decade mainly thanks to the reactor and accelerator-based experiments~\cite{An:2016ses,Abe:2017vif}. Nowadays, the neutrino oscillation parameters have been measured with certain precision~\cite{deSalas:2017kay,nufit}, except for the Dirac phase encoding the possibility that leptons violate the charge-parity (CP) symmetry. In this so-called {\it three active neutrino framework}, the neutrino mass ordering, whether the third mass eigenstate is the upper (normal ordering) or the lower (inverted ordering) of the three states, is also unknown. Future neutrino oscillation experiments are expected to resolve both important missing pieces and also to improve over the current precision of the neutrino oscillation parameters. In particular, there is a quest for establishing if the atmospheric mixing angle is maximal, and if not, what would be its correct octant. Besides providing information on the unknowns, in the precision era, new physics signals might emerge as subleading effects of the three neutrino paradigm or as a new oscillation phase(s). This last scenario is mainly motivated by results of short-baseline experiments~\cite{Aguilar:2001ty,AguilarArevalo:2007it,AguilarArevalo:2010wv,Aguilar-Arevalo:2018gpe} which call for a new neutrino flavor state that has to be sterile, i.e. it can not interact with the Standard Model gauge bosons. So far, there is no indication of a new oscillation phase and running experiments have constrained a large part of the parameter space, at least in the economical $3+1$ oscillation framework~\cite{Abe:2014gda,Adamson:2011ku,TheIceCube:2016oqi,An:2016luf,MINOS:2016viw,Adamson:2016jku,Aartsen:2017bap,Adamson:2017zcg}. Several efforts are devoted to discover a sterile oscillation at the eV mass scale or to completely rule out this hypothesis. For instance, at FERMILAB, there is a Short-Baseline Neutrino Oscillation Program  (SBN)~\cite{Antonello:2015lea}, which is expected to provide a definitive answer to this matter. However, there are several beyond the standard three-neutrino oscillation scenarios, which might be considered as a the subleading effect, that can be probed in future long and short-baseline neutrino experiments. Here we focus on Large Extra Dimensions (LED) and the possibility that its signals be differentiated from the sterile hypothesis at the SBN program. Other proposals can be tested in SBN facility for instance the search for multiple sterile states~\cite{Fan:2012ca,Khruschov:2016sjl,Cianci:2017okw} and  MeV-scale sterile decay~\cite{Gninenko:2010pr,Dib:2011jh,Ballett:2016opr}.

Initially, the main motivation for introducing extra space-time dimensions was to lower high energy scales, as for instance the GUT~\cite{Dienes:1998vh,Dienes:1998vg} or the Planck scale, even to the TeV energy scale~\cite{ArkaniHamed:1998rs,ArkaniHamed:1998nn,Antoniadis:1998ig}. This appeared as an alternative to the usual seesaw mechanism that in its natural form calls for a high energy scale to suppress the active neutrino masses. Since right-handed neutrinos are singlets under the Standard Model~(SM) gauge group, they are one of the candidates that can experience extra space-time dimensions and therefore collect an infinite number of Kaluza-Klein excitations~\cite{Dienes:1998sb,Barbieri:2000mg}. The other SM fermions are restricted to a brane and therefore experiencing four dimensions only. In this way, the Yukawa couplings between the right-handed neutrinos and the active ones are suppressed by the volume factor after compactification of the extra dimensions. In this context, neutrinos acquire a Dirac mass that is naturally small, however, other alternatives violating lepton number are possible~\cite{Dienes:1998sb}. It is phenomenological appealing to  considered an asymmetric case where one of the extra dimensions is `large' respect to the others, effectively reducing the problem to be five dimensional~\cite{Barbieri:2000mg,Dienes:1998sb,Davoudiasl:2002fq}. In this letter,  we consider the model for Large Extra Dimensions (LED) from Ref.~\cite{Davoudiasl:2002fq} (which is based on previous works in Refs.~\cite{Dienes:1998sb,Barbieri:2000mg,Mohapatra:1999zd}), and recently considered in the context of DUNE in Ref.~\cite{Berryman:2016szd}, with three bulk neutrinos (experiencing extra space-time dimensions) coupled to the three active brane neutrinos.

In this letter, we consider neutrino oscillations within the LED model with three bulk neutrinos coupled to the three active brane neutrinos, which effectively act as a large number of sterile neutrinos in contrast to the usual oscillation of light sterile neutrinos at the eV energy scale. Our goal is to establish the sensitivity of the SBN program to neutrino oscillations in the LED model. This letter is organized in the following way. We first introduce the LED formalism in Section~\ref{formalism}. The SBN program and the experimental details used in our numerical simulations are condensed in Section~\ref{simulation}. Our results are presented in Section~\ref{results}. Finally, we conclude and summarize in Section~\ref{summary}.

\section{Formalism} 
\label{formalism}

In general, it is assumed the right-handed neutrino (bulk fermions~\cite{Davoudiasl:2002fq}) can propagate in more than four dimensions while the left-handed neutrino $\nu_{L}$, and the SM Higgs $H$, are confined to the four-dimensional  brane. It is also assumed that one of the extra space-time dimensions is larger than the others so that effectively it is enough to consider five dimensions in total. A Dirac fermion $\Psi^\alpha$ in five dimensions can be decomposed into two component spinors (Weyl fermions), $\psi_L$ and $\psi_R$ and after the extra dimension is compactified a {\it natural} coupling with $\nu_L$ emerges~\cite{Dienes:1998sb} and, as a result, Dirac neutrino masses are obtained~\cite{Dienes:1998sb,Barbieri:2000mg,Mohapatra:1999zd,Davoudiasl:2002fq}. Along this letter we follow the model with three bulk neutrinos coupled via Yukawa couplings to the three active brane neutrinos, the  so-called $(3,3)$ model in Ref.~\cite{Davoudiasl:2002fq}.  Other formulations for large extra dimension models are possible as described in Ref.~\cite{Carena:2017qhd}.

The action in the $(3,3)$ model is given by:

\begin{equation}\label{eq:action}
S=\int d^4x dy \,\bar{\Psi}^\alpha \Gamma_A\, i\partial^A \Psi^\alpha+\int d^4x\left[\bar{\nu}^\alpha_L \gamma_\mu i\partial^\mu \nu_L^\alpha+\lambda_{\alpha \beta} H\, \bar{\nu}^\alpha_L\, \psi^\beta_R(x,0)+ \text{H.c.} \right]
\end{equation}
where $y$ is the coordinate of the extra compactified dimension, $\Gamma_A$ are the five-dimensional Dirac matrices for $A=0,...,4$, and $\lambda_{\alpha \beta}$ the Yukawa couplings. To compactify the action in Eq.~(\ref{eq:action}) one need to expand the the five-dimensional Weyl fields $\psi_{L,R}$ in Kaluza-Klein (KK) modes $\psi^{(n)}_{L,R}$ (with $n=0, \pm 1,...,\pm \infty$) and also to impose suitable periodic boundary conditions~\cite{Dienes:1998sb}. It is convenient to define the following linear combinations:

\begin{equation}
\begin{split}
\nu^{\alpha(n)}_R&=\frac{1}{\sqrt{2}}\left(\psi^{\alpha (n)}_R+\psi^{\alpha(-n)}_R\right)\\
\nu^{\alpha(n)}_L&=\frac{1}{\sqrt{2}}\left(\psi^{\alpha (n)}_L+\psi^{\alpha(-n)}_L\right)\,,
\end{split}
\end{equation}
for $n>0$, and also $\nu^{\alpha(0)}_R\equiv \psi^{\alpha (0)}_R$. Therefore, after electroweak symmetry breaking, the Lagrangian mass terms that results from Eq.~(\ref{eq:action}) are given by:

\begin{equation}\label{eq:lag}
\mathcal{L}_{\text{mass}}= m^D_{\alpha  \beta}\left(\bar{\nu}^{\alpha(0)}_R \nu^{\beta}_L+\sqrt{2}\sum_{n=1}^\infty \bar{\nu}^{\alpha(n)}_R \nu^{\beta}_L\right) + \sum_{n=1}^\infty \frac{n}{R_{\text{LED}}}\bar{\nu}^{\alpha(n)}_R \nu^{\beta(n)}_L + \text{H.c.}\,,
\end{equation}

Where $m^D$ is the Dirac mass matrix that is proportional to the Yukawa couplings and can be written in terms of the fundamental mass scales of the theory~\cite{Dienes:1998sb,Davoudiasl:2002fq}, and $R_{\text{LED}}$ is the compactification radius. It is useful to consider a basis in which the Dirac mass is diagonal~\cite{Davoudiasl:2002fq} $r^\dagger \,m^D\,l=\text{diag}\{m_i^D\}$, by defining pseudo mass eigenstates $\mathcal{N}^i_{L,R}=\left(\nu^{i(0)},\nu^{i(1)},\nu^{i(2)},...\right)^T_{L,R}$~\cite{Berryman:2015nua}, such that the mass Lagrangian in Eq.~(\ref{eq:lag}) can be written $\mathcal{L}_{\text{mass}}=\sum_{i=1}^{3} \bar{\mathcal{N}}^i_R\, M^i\,\mathcal{N}^i_L+\text{H.c.}$ where $M^i$ is the infinite-dimensional matrix given by~\cite{Barbieri:2000mg,Davoudiasl:2002fq}:

\begin{eqnarray}\label{eq:mass}
M^i=
\begin{pmatrix}
m_i^D&0&0&0&\ldots\\
\sqrt{2}m_i^D&1/R_{\rm{ED}}&0&0&\ldots\\
\sqrt{2}m_i^D&0&2/R_{\rm{ED}}&0&\ldots\\
\vdots&\vdots&\vdots&\vdots&\ddots
\end{pmatrix}.
\end{eqnarray}
To find the neutrino masses and the relevant unitary matrices $L_i(R_i)$ that relate the mass eigenstates $\mathcal{N}^\prime_{iL(R)}$ with the pseudo eigenstates $\mathcal{N}_{iL(R)}$, $\mathcal{N}^\prime_{iL(R)}=L(R)_i^\dagger \mathcal{N}_{iL(R)}$, one needs to perform the bi-diagonalization $R^\dagger_i\,M_i\,L$. However, since we are mostly interested in the relation of the active brane neutrino states $\nu_L^\alpha$ with the mass eigenstates, it is enough to consider only the left matrices $l$ and $L_i$. $L_i$ is obtained from the diagonalization of the Hermitian matrix $M_i^\dagger M_i$ while $l$ is the unitary $3\times 3$ matrix involved in the $m^D$ diagonalization.

Effectively the active neutrino flavor states, can be finally written in terms of the mass eigenstates (as composed of the KK $n$-modes of the fermion field), as follows:
\begin{equation}\label{nustate}
\nu_{\alpha\,L}=\sum_{i=1}^3 
 l_{\alpha i} \nu_{i\,L}^{(0)}=\sum_{i=1}^3 l_{\alpha i} \,\sum_{n=0}^\infty L_i^{0n} \nu_{i\,L}^{\prime (n)} \equiv \sum_{i=1}^3   \sum_{n=0}^\infty W_{\alpha i}^{(n)} \nu_{i\,L}^{\prime (n)} 
 \,.
\end{equation}
 where $W_{\alpha i}^{(n)}$ is the amplitude in the LED case. We recover the usual three-neutrino case when $W_{\alpha i}^{(n)} \to  l_{\alpha i}$.

Formally, the mass eigenvalues and the eigenvectors of $M_i$ in Eq.~(\ref{eq:mass}) are obtained from the diagonalization of the matrix $R_{\rm ED}^2\, M_i^\dagger M_i$ by assuming a maximum integer value for the KK-modes $k_{\text max}$ and then taking the limit $k_{\text max}\to \infty$~\cite{Dienes:1998sb,Barbieri:2000mg}.
The $L_i^{0n}$ matrix in Eq.~(\ref{nustate}) is explicitly given by:
\begin{equation}\label{smatrix}
\left(L_i^{0n}\right)^2=\frac{2}{1+\pi^2\left(R_{\rm ED}m_i^D\right)^2+\left[\lambda_i^{(n)}/\left(R_{\rm ED}m_i^D\right)\right]^2}\,,
\end{equation}
where the neutrino mass eigenstates are equal to $\lambda_i^{(n)}/R_{ED}$ and therefore each one of them is composed of $n$-KK modes. $\lambda_i^{(n)}$ in Eq.~(\ref{smatrix}) corresponds to the eigenvalues of the full $n\times n$ neutrino mass matrix and can be calculated from the following transcendental equation:
\begin{equation}\label{lambdas}
\lambda_i^{(n)}-\pi\left(R_{\rm ED}\,m_i^D\right)^2 \text{cot}{\left(\pi \lambda_i^{(n)}\right)}=0\,,
\end{equation}
and the roots $\lambda_i^{(n)}$ are constrained such that they belong to the range $[n,n+1/2]$~\cite{Dienes:1998sb}. In order to make a physical sense of the formalism, one should assume that the most active state is obtained for $n=0$. Additionally, if we go to the limit $R_{\rm ED}\,m_{i}^D\ll 1$ then $\lambda^{(0)}\to R_{\rm ED}\,m_{i}^D$, and following Eq.~(\ref{smatrix}) $L_{i}^{00} \to 1$, therefore recovering the standard result where $l_{\alpha i}\to U_{\alpha i}$ is the lepton mixing matrix that is usually parametrized by three rotations~\footnote{The three rotations are in general complex, accounting for the three physical CP phases. However, neutrino oscillations are insensitive to the two Majorana phases, and therefore, only sensitive to the Dirac CP phase. In this case the more used parametrization is written as two real rotations plus a complex one.}, through the three mixing angles $\theta_{ij}$, and the Dirac CP phase $\delta$.

Assuming the mostly active mass state is related with the lightest mass state in the KK-tower implies a relation between the eigenvalues of this LED framework, obtained by Eq.~(\ref{lambdas}), with the square mass differences obtained in the three-neutrino case. This relation can be written as:
\begin{equation}\label{numass}
\frac{\left(\lambda_k^{(0)} \right)^2-\left(\lambda_1^{(0)} \right)^2}{R_{\rm ED}^2}=\Delta m_{k1}^2
\end{equation}
with $\Delta m_{k1}^2$ is the solar ($k=2$) and the atmospheric ($k=3$) squared mass differences. Therefore, the existing values on the squared mass differences of the active neutrino mass eigenstates $\Delta m_{k1}^2$, Ref.~\cite{Esteban:2016qun,nufit}, constrain the parameter space $(m_{i}^D, R^{-1}_{\rm ED})$ of the LED model. Thus, a good strategy is to use this information before scanning the parameter space. Basically, $\lambda_i^{(0)}$, i=1,2,3  are fixed by the $m_i^D$ in Eq.~(\ref{lambdas}), and using Eq.~(\ref{numass}) for k=2,3 we got  a constrain between $m_1^D$, $m_2^D$ and $m_3^D$~\cite{BastoGonzalez:2012me}. With these constraints, we have now only two independent parameters $m_1^D$ and $R_{\rm ED}$ that we will rename from now on as  $m_1^D \to m_0$ for normal mass ordering.  Similarly, one can follow the same procedure for the inverted mass ordering, and this case the two independent parameters are  $m_3^D \to m_0$ and $R_{\rm ED}$. In the cases where the condition in Eq.~(\ref{numass}) is not fulfilled by the $(m_{1}^D, R^{-1}_{\rm ED})$ combination, we quoted the excluded region as {\em excluded by squared mass differences constraints}. We will comeback to this point in Section~\ref{simulation}.

In the LED framework the neutrino mixing matrix $W$, as defined in Eq.~(\ref{nustate}), is in general different to the standard three neutrino mixing matrix $U$. To avoid spoiling the neutrino oscillations observations, condensed in part as constraints on the mixing angles $\theta_{ij}$ ($i,j = 1,2,3$) in {\em scenario of three-neutrino scheme} (with values in Ref.~\cite{deSalas:2017kay,Esteban:2016qun,nufit}), the mixing angles in the LED framework have to be redefine. Following the procedure from Ref.~\cite{Berryman:2016szd} we have defined {\em new} mixing angles $\phi_{ij}$ ($i,j = 1,2,3$) in the LED scenario such that the lowest mass state in KK tower, $n=0$,  have the $W_{\alpha i}^{(0)}$ amplitude equal to the numerical value of $U_{\alpha i}$: $U_{\alpha i}=W_{\alpha i}^{(0)}= l_{\alpha i}  L_i^{00}$. From this relation we can get the mixing angles in the LED framework,  $\phi_{ij}$, related with the solar and atmospheric mixing angles,  $\theta_{ij}$. Explicitly we have  using the elements of mixing matrix $|U_{e2}|$, $|U_{e3}|$ and $|U_{\mu 3}|$
\begin{eqnarray}
 & & \sin\phi_{13} = \frac{\sin \theta_{13}}{\left( L^{00}_{3} \right)}\quad\quad  
\cos \phi_{13} \sin \phi_{12} = \dfrac{\cos \theta_{13} \sin \theta_{12}}{\left( L^{00}_{2} \right)} \nonumber \\
& & \cos \phi_{13} \sin \phi_{23} = \dfrac{\cos \theta_{13} \sin \theta_{23}}{\left( L^{00}_{3} \right)}\,. 
\label{w-rename}
\end{eqnarray}
From now on, the mixing angles $\phi_{ij}$ in the LED formalism are given by the values in Eq.~(\ref{w-rename}). For some values of $m^{D}_{i}$ and $R_{\rm{ED}}$ the $L^{00}_{i}$ value can be smaller than the numerator in Eq.~(\ref{w-rename}) such that  $\sin\phi_{ij}>1$ and thus unphysical. In this way, values of $m^{D}_{i}$ and $R_{\rm{ED}}$ that results in this unphysical $\phi_{ij}$ will be disregarded and we have quoted them as {\em excluded by mixing angle constraints}. We will comeback to this point in Section~\ref{simulation}.

\section{Simulation} 
\label{simulation}

In this section, we describe the experimental set-up and our working assumptions that we followed in the sensitivity analyses presented in Section~\ref{results}. The SBN experimental proposal will align three liquid argon detectors in the central axis of the Booster Neutrino Beam (BNB), located at FERMILAB~\cite{Antonello:2015lea}. Table~\ref{tab:detinfo} gives the SBN detector names, active masses, locations, and protons on target POT. We computed the expected number of events of SBN facility by implementing the detectors in the GLoBES~\cite{Huber:2004ka, Huber:2007ji} c-library, following the proposal description. The flux information for both neutrino and anti-neutrino modes was taken from Ref.~\cite{Admas:2013xka}, and the neutrino-argon cross section was taken from inputs to GLoBES prepared for Deep Underground Neutrino Experiment (DUNE) simulation~\cite{Acciarri:2015uup}, with the cross section inputs, originally generated using GENIE 2.8.4~\cite{Andreopoulos:2009rq}. 

\begin{table}[t!]
\centering
\begin{adjustbox}{width=1\textwidth}
\begin{tabular}{||c c c c ||} 
 \hline
 Detector & Active Mass & Distance from BNB target & POT \\ [0.5ex] 
 \hline\hline
Lar1-ND & 112 t & 110 m &  $6.6\times 10^{20}$ \\ 
MicroBooNE & 89 t & 470 m & $1.32\times 10^{21}$  \\
ICARUS-T600 & 476 t & 600 m & $6.6\times 10^{20}$ \\ [1ex] 
 \hline
  \hline
  \multicolumn{2}{|c}{\textbf{Electron Neutrino Appearance Channel}} &  \multicolumn{2}{c|}{ \textbf{Muon Neutrino Disappearance Channel}}\\ [0.5ex] 
 \hline\hline
Energy Bin Size (GeV) & Energy Range (GeV) &  Energy Bin Size (GeV) & Energy Range (GeV) \\
\hline 
0.15 & 0.2-1.10 & 0.10 & 0.2-0.4 \\
0.20 & 1.10-1.50 & 0.05 & 0.4-1.0 \\
0.25 & 1.50-2.00 & 0.25 & 1.0-1.5 \\
1.00 & 2.00-3.00 & 0.50 & 1.5-3.0 \\
\hline
\end{tabular}
 \end{adjustbox}
\caption{Upper: SBN detector active masses and distances from local of neutrino production. Lower: Energy range  and energy bin size of the electron and muon sample used in this analysis.}
\label{tab:detinfo}
\end{table}

The SBN facility will search for oscillations in two channels: 1) electron neutrino appearance from muon neutrino conversion ($\nu_{\mu} \rightarrow \nu_{e}$) and 2) muon neutrino disappearance ($\nu_{\mu} \rightarrow \nu_{\mu}$) from muon neutrino survival. We considered a Gaussian detector energy resolution function with a width of $\sigma (E) = 6\%/\sqrt{E [\rm{GeV}]}$ for muons and $\sigma (E) = 15\%/\sqrt{E [\rm{GeV}]}$ for electrons, according to Ref.~\cite{Cianci:2017okw}. The energy range for the neutrino event reconstruction extends from 0.2 GeV to 3 GeV where each channel has different bin widths, as described in the Table~\ref{tab:detinfo}. 
We simulated three years of operation for the neutrino beam in Lar1-ND and ICARUS-T600 detectors and six years in MicroBooNE detector. It is important to emphasize that the detectors do not make a distinction between neutrinos and anti-neutrinos, so neutrinos and anti-neutrinos  events are added in our simulations. After event reconstruction, we included an efficiency factor for each channel in order to mimic event rates from collaboration proposal~\cite{Antonello:2015lea}.  

In the presence of LED, the relations in Eq.~(\ref{numass}) and Eq.~(\ref{w-rename}) gives the squared mass differences and the mixing angles in terms of the standard oscillation parameters. When simulating neutrino event rates, to perform the different studies along this letter, we used the best-fit values for the oscillation parameters in the standard three-neutrino framework presented in Nu-Fit 3.2 (2018)~\cite{Esteban:2016qun,nufit}. The LED parameters are the lightest neutrino mass $m_{0}$ (for normal ordering $m_{0} = m_{1}^{D}$ while for inverted ordering $m_{0} = m_{3}^{D}$) and the radius of extra dimension $R_{\rm{ED}}$.

\begin{figure}[t]
\centering
\includegraphics[scale=0.34]{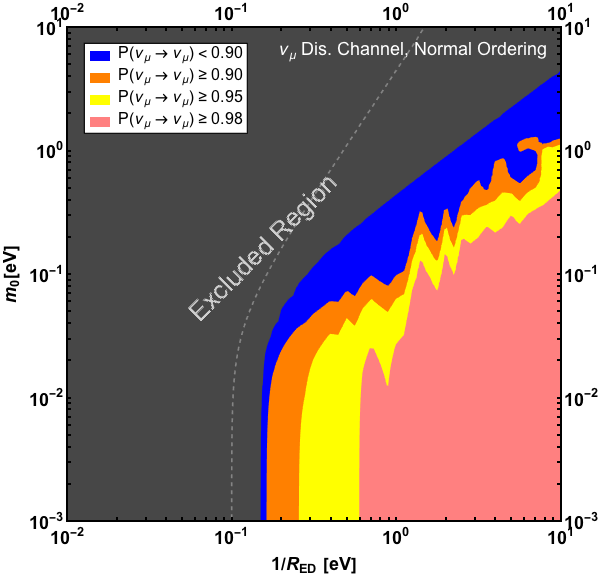}\hfill
\includegraphics[scale=0.34]{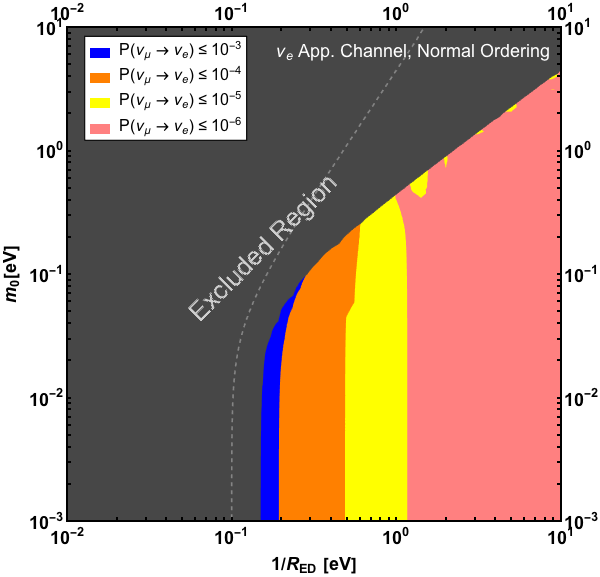}
\includegraphics[scale=0.34]{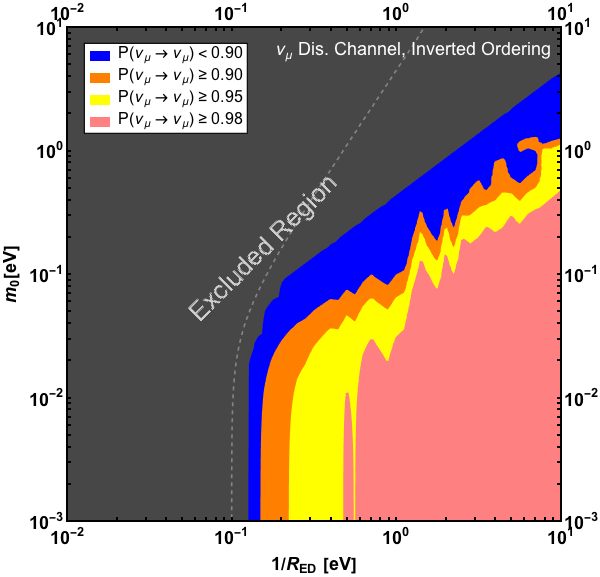}\hfill
\includegraphics[scale=0.34]{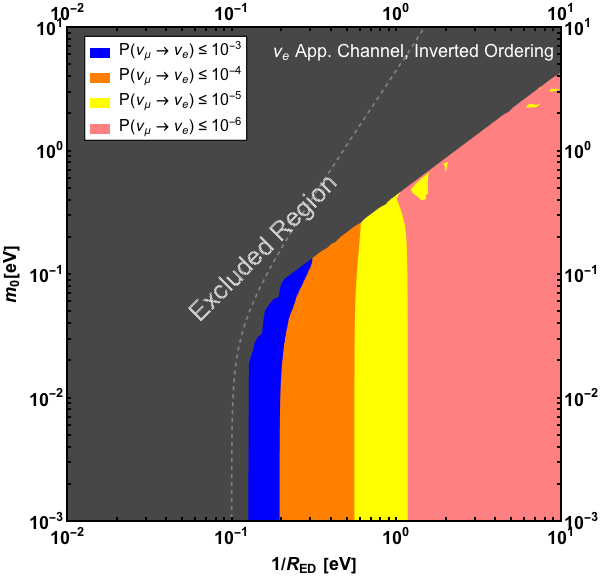}
\caption{Iso-probability regions for different values of LED parameters, $m_{0}$ and $R_{\rm ED}$. In the left (right) panels we
have  $\nu_{\mu} \rightarrow \nu_{\mu}$ ($\nu_{\mu} \rightarrow \nu_{e}$). In the top (bottom) we show the normal (inverted) ordering.
We chose here a typical short-baseline $L/E_{\nu}$ of 1.2 km/GeV, see text for details, and we compute probabilities using the first 40 KK modes. The gray shaded region is excluded due to neutrino oscillation data (see Sec.\ref{formalism}).}
\label{fig:prob}
\end{figure}

In Figure~\ref{fig:prob} the behavior of the oscillation probability for different $m_{0}$ and $R_{\rm{ED}}$ values is shown, considering an $L/E_{\nu}$ of $1.2$~km/GeV in both appearance and disappearance channels for both normal and inverted neutrino mass ordering. The $L/E_{\nu}$ value was calculated using the ICARUS baseline $L = 0.6$~km and the energy $E_{\nu} = 0.5$~GeV, which corresponds to the region in the neutrino energy spectrum where most of the events are expected~\cite{Antonello:2015lea}. We noticed that for all LED parameters in the  $R_{\rm{ED}}^{-1}-m_{0}$ plane, the appearance probability is not larger than 10$^{-3}$ and almost all survival probability is larger than 0.9. The gray shaded region is excluded by neutrino oscillation data, with the relations Eq.~(\ref{numass}) and Eq.~(\ref{w-rename}), as described in Section~\ref{formalism}.

In the following, we assume {\em forward horn current} (FHC) beam mode and we defined signal and background for each one of the SBN oscillation channels as follows:

\begin{itemize}
\item \textbf{Muon neutrino disappearance channel}:
\begin{enumerate}
\item \textit{Signal:} Survival of muon neutrinos ($\nu_{\mu} \rightarrow \nu_{\mu}$) from the beam which interact with liquid argon through weak charged-current (CC) producing muons in the detectors.
\item \textit{Background}: The only background contribution considered by the collaboration comes from neutral-current (NC) charged pion production, where the pion produced in the BNB target interacts with argon and can be mistaken for a muon~\cite{Antonello:2015lea}. This contribution is small due to the track cutting imposed in the event selections and we did not consider it in our simulations.  
\end{enumerate}

\item \textbf{Electron neutrino appearance channel}:

\begin{enumerate}
\item \textit{Signal:} electron neutrinos coming from muon neutrino conversion ($\nu_{\mu} \rightarrow \nu_{e}$) which interacts through CC producing electrons in the detectors.

\item \textit{Background}: The main background contribution comes from the survival of intrinsic electron neutrinos ($\nu_e \rightarrow \nu_e$) in the beam, beam contamination. We also considered muons (muon neutrinos from the CC interaction), which can be mistaken for electrons. NC photon emission, cosmic particles and dirty events were not considered in our simulation, which corresponds to a background reduction of $8.4\%$ for Lar1-ND, $14\%$ for MicroBooNE and $13\%$ for ICARUS-T600, respect to the total number of background events expected by the collaboration in the electron neutrino channel~\cite{Antonello:2015lea}. 
\end{enumerate}
\end{itemize}

The information on the neutrino fluxes, neutrino cross section, energy resolution of leptons and backgrounds used in the analysis were compiled using the AEDL format (to be used with the GLOBES c-library), in order to perform the different sensitivity analyses of SBN program at FERMILAB. These files are available under request following Ref.~\cite{aedl-sbn}.

Since one of the main goals of the SBN program is to detect or rule out sterile neutrino oscillations, we introduce the generalities of the $3+1$ case right now. Later, we will not only take it as a reference but also we will quantify the discrimination power of the SBN program between the two models, the $3+1$ and the LED. Several neutrino experiments have performed a sensitivity analysis in the specific scenario of the so-called 3+1 model, where one sterile neutrino is added to the three active neutrino framework. In this 3+1 framework, active and sterile neutrinos mix and three new oscillation frequencies appear, thanks to the four mass eigenstates, which can be written in terms of only $\Delta m_{41}^2$, the solar, and the atmospheric splittings. The additional mass eigenstate is the source of short-baseline oscillations mainly driven by the square mass difference $\Delta m_{41}^2$, and the effective amplitudes $\sin^2 2\theta_{\mu e} \equiv 4|U_{e 4}|^{2}|U_{\mu 4}|^{2}$ and $\sin^2 2 \theta_{\mu \mu} \equiv 4 \left( 1 - |U_{\mu 4}|^{2} \right) |U_{\mu 4}|^{2}$ defined by the elements of the $4\times 4$ lepton mixing matrix. We have successfully reproduced the results of the SBN experimental proposal regarding the sensitivities to the sterile parameters by the implementation of the muon disappearance and electron appearance oscillation channels. These sensitivities will be considered and shown in Section~\ref{results}.

In the following sections, we present results based on different sensitivity analysis, using both muon and electron appearance channels, unless otherwise stated. We studied three cases assuming a given event energy spectrum for `data' (or `true' events) and we have performed a hypothesis testing based on a Poisson $\chi^2$ function for the different models:
\textbf{1)} `data' simulated assuming an energy spectrum defined by the three-neutrino case an testing the LED hypothesis, i.e., the usual sensitivity analysis, \textbf{2)} `data' simulated assuming an energy spectrum distributed with the LED model and testing the standard oscillation scenario. Here we investigated the SBN potential of measuring the LED parameters $R_{\rm{ED}}$ and $m_{0}$. Finally, \textbf{3)} `data' simulated assuming an energy spectrum distributed with the 3+1 model, where we evaluated the discrimination power of SBN to distinguish LED hypothesis from other models accommodating light sterile neutrino oscillations. We also performed sensitivity calculations for the 3+1 model in appearance and disappearance channels in order to explore relations between LED and 3+1 signatures. The results are shown in the next section.

\section{Results}\label{results}

For the sensitivity analysis, total normalization errors in signal and background were set to $10\%$, and all parameters that were not shown in the plots were fixed to their best-fit values. We tested that our sensitivity results are independent of the $\delta_{\rm{CP}}$ value. For simplicity, we set $\delta_{\rm{CP}} = 234^{o}$ for normal ordering and $\delta_{\rm{CP}} = 278^{o}$ for inverted ordering, according to Ref.~\cite{nufit}. 

\begin{figure}[t]
\centering
\includegraphics[scale=0.35]{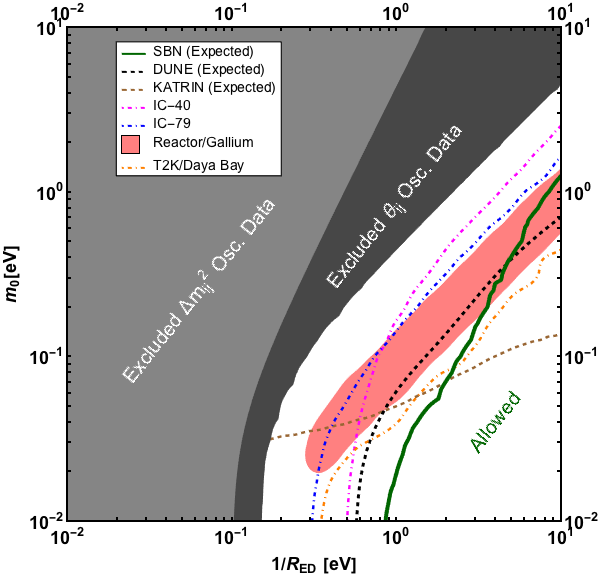}
\includegraphics[scale=0.35]{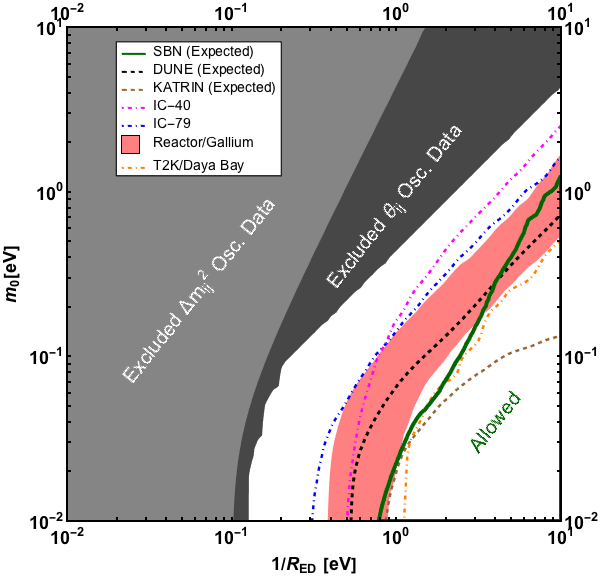}
\caption{Left (Right) panel: Sensitivity limits for the LED parameters, $R_{\rm{ED}}$ and $m_{0}$, considering normal (inverted) ordering of neutrino masses. The regions for LED sensitivity, considering both channels, muon disappearance and electron appearance channels, are to the top-left of the curves. Here, we show our 90\% C. L. line from SBN limit (green), the 95\% C. L. lines from DUNE (black)~\cite{Berryman:2016szd}, ICECUBE-40 (magenta) and ICECUBE79 (blue)~\cite{Esmaili:2014esa}, and 95\% C. L. from combined analysis of T2K and Daya Bay (gold)~\cite{DiIura:2014csa}. The 90\% C. L. line from KATRIN sensitivity analysis is also shown (brown)~\cite{BastoGonzalez:2012me} and the pink regions are preferred at 95\% C. L. by the reactor and Gallium anomaly~\cite{Machado:2011kt}. The light and dark gray regions are excluded due to neutrino oscillation data.}
\label{fig:LED-all}
\end{figure}

Figure~\ref{fig:LED-all} shows SBN sensitivity limit with $90\%$ of confidence level (C.L.) in the green curve for normal (left panel) and inverted ordering (right panel), compared with other limits: Sensitivity limits at 95 \% of C. L. for DUNE experiment (black-dashed curve) presented in Ref.~\cite{Berryman:2016szd}, as well as ICECUBE-40 data and ICECUBE-79 data (dot-dashed magenta and blue curves, respectively) from Ref.~\cite{Esmaili:2014esa}, and the combined analysis of T2K and Daya Bay data (dot-dashed gold curve) presented in Ref.~\cite{DiIura:2014csa} are shown. The preferred region (in pink) at 95\% C. L. by Gallium and Reactor anti-neutrino experiments from the analysis in Ref.~\cite{Machado:2011kt} is also included. Finally, sensitivity limits for KATRIN at $90\%$~C. L. (dashed brown curve) due to kinematic limits in beta decay estimated in Ref.~\cite{BastoGonzalez:2012me} are shown. The gray shaded regions are the parameters excluded by measurements of square mass differences $\Delta m_{\rm{sol}}^{2}$ and $\Delta m_{\rm{atm}}^{2}$ (light gray) and of mixing angles $\theta_{12}$, $\theta_{13}$ and $\theta_{23}$ (dark gray). It is important to mention that excluded region due to mixing angle measurements also covers excluded region due to square mass differences. An additional constrain to the LED parameters comes from MINOS analysis in Ref.~\cite{Adamson:2016yvy} where a similar restriction curve to the one from ICECUBE was obtained. When $m_{0} \rightarrow 0$, MINOS constrains $R_{\rm{ED}} < 0.45\,\mu$m (or $R_{\rm ED}^{-1}>0.44$~eV) for normal ordering.    

We can see that the SBN program is sensitive to the LED parameters and this sensitivity is very competitive, respect to other facilities shown in the plot. This happens specifically for the lower $m_0$ region and particularly for normal ordering. Comparing with the constraints from other experiments, the SBN sensitivity for LED mechanism is the better than any other constraints in the region when $m_0 <2\times 10^{-1}$ for normal ordering, and in this region, the maximum sensitivity of our analysis for $R_{\rm{ ED}}$ is better than any other oscillation experiment which we trace to the fact that we are testing LED in a short-baseline experiment for the first time, all other sensitivity results corresponds to long-baseline experiments. With respect to the {\em reactor anomaly} allowed region, the SBN program has the potential to ruled out completely this anomaly for any value of $m_0 < 2\times 10^{-1}$. For higher values of $m_0$, the DUNE experiment~\cite{Berryman:2016szd} have the potential to exclude the reactor anomaly allowed region, complementing SBN.

\subsection{Sensitivity to a non-zero LED oscillation effect on SBN}
\label{positive}

\begin{figure}[t]  
\centering
\includegraphics[scale=0.35]{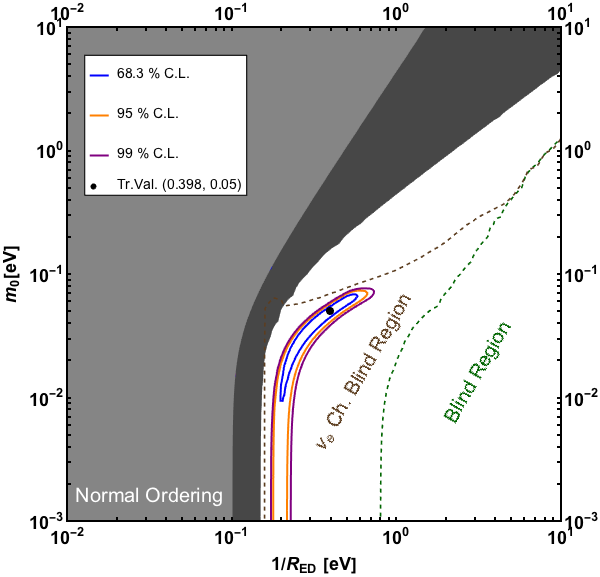}
\includegraphics[scale=0.35]{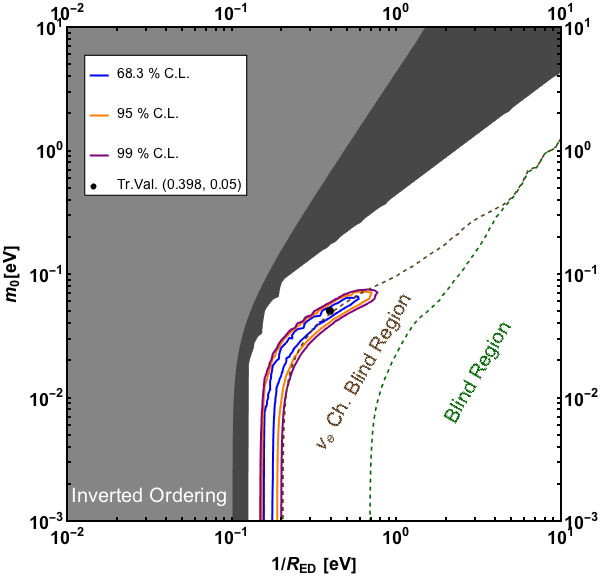}
\caption{ Left (right) panel: Allowed regions for the `true' LED parameters $m_{0} =0.05 \ \text{eV}$ and $1/R_{\text{ED}} = 0.398 \ \text{eV}$ and assuming as test model the LED scenario for normal (inverted) ordering. All the other oscillation parameters were fixed to their best fit values. The dashed green (dashed brown) curve show respectively the SBN sensitivity to the both  muon disappearance and electron appearance channel (only electron appearance channel). The region denoted by {\em Blind region} ({\em $\nu_e$ Ch. Blind region}) is the region were are not sensitive respectively to the muon neutrino  disappearance (electron neutrino appearance).}
\label{fig:LED-fit}
\end{figure}

In order to investigate the potential of SBN to measure the LED parameters, neutrino events were calculated in the same fashion than for the previous sensitivity analysis, but assuming now the LED model with $m_{0}=0.05\,\text{eV}$ and $1/R_{\text{ED}} =0.398\,\text{eV}$ as the `true' values, and testing the LED scenario. All the standard oscillation parameters (which are included in the LED parameters) were fixed to their best-fit values from Refs.~\cite{Esteban:2016qun,nufit} as described in Section~\ref{formalism}. Figure~\ref{fig:LED-fit} shows the allowed regions consistent with the computed events with the true value (black dot) at $68.3\%$ of C.L. (blue curve), $95\%$ of C.L. (orange curve) and $99\%$ of C.L. (purple curve) for both normal ordering (left panel) and inverted ordering (right panel). 

We also included in Figure~\ref{fig:LED-fit} the sensitivity result obtained in Figure~\ref{fig:LED-all} (dashed green line), which we called {\em Blind Region}, i.e., the region that agrees with the standard three-neutrino scenario, being in this way, `blind' to LED effects. Any point inside the {\em Blind Region} will have a null result either for the muon disappearance channel or for the electron neutrino appearance channel. The {\em $\nu_e$ Ch. Blind Region} presented in Figure~\ref{fig:LED-fit} (dashed brown line) is the result of the sensitivity analysis performed {\em only} with the computed events from electron neutrino appearance channel. Any point inside the {\em $\nu_e$ Ch. Blind Region} will have a null result for the electron neutrino appearance channel.
The `true' LED parameters were chosen around the  $\nu_e$ Ch. Blind Region, but outside the Blind Region for both mass ordering.

It is worth noticing that since the electron neutrino appearance probability is smaller than $10^{-3}$ for LED, as shown in Figure~\ref{fig:prob}, one might not expect a sensitivity exclusion limit from the appearance channel, i.e., all the obtained sensitivity is shown in Figure~\ref{fig:LED-all} would come from the muon disappearance channel. However, when we computed the sensitivity curve only considering electron appearance channel, we obtained the exclusion limit showed in Figure~\ref{fig:LED-fit} (dashed brown line). In fact, we have a sensitivity curve from electron appearance channel when we consider changes in background profile due to LED effects. The electron neutrino survival probability induced by the LED parameters decreases the intrinsic electron neutrinos from the beam, which is the majority contribution to our background. In other words, we have sensitivity due to the decrease in the number of backgrounds and not by the increase in the signal. A similar effect was found in Ref.~\cite{Cianci:2017okw}.

Although not shown in Figure~\ref{fig:LED-fit}, we repeated the same analysis with other LED {\em true values} located inside the exclusion region for both electron and muon neutrino channels (outside the {\em Blind Region} and the {\em $\nu_e$ Ch. Blind Region}). In this case, we have a non-null result in {\em both} muon disappearance and electron neutrino appearance channels, and therefore the LED parameters that explain this results are unique. As a consequence of this, and due to  the logarithmic scale in the plot, we obtained small and concentrated regions around the chosen `true' values, which results in a precision of SBN experiment to the LED parameters below 1\%.

\subsection{3+1 scenario at SBN: sensitivity and accuracy of the measurement}
\label{3p1}

In the  standard three-neutrino scenario, we expect no oscillations in SBN due to its short-baseline and the energies considered. Now, if SBN `sees' an oscillation, it will corresponds to a beyond the standard three-neutrino scenario signal that might be interpreted as an sterile neutrino oscillation. In the 3+1 scenario, the neutrino probabilities for short-baseline distances are given by~\cite{Conrad:2016sve}:

\begin{align}
P^{3+1}_{\nu_{\mu} \rightarrow \nu_{e}} &= \sin^2 2 \theta_{\mu e} \sin^2 \left( \Delta m^{2}_{41} L/(4 E_{\nu})  \right) 
\label{eq:tconv31} \\
P^{3+1}_{\nu_{\mu} \rightarrow \nu_{\mu}} &= 1 - \sin^2 2 \theta_{\mu \mu} \sin^2 \left(\Delta m^{2}_{41} L/(4 E_{\nu}) \right)\,, \label{eq:tsurv31}  \\
 P^{3+1}_{\nu_{e} \rightarrow \nu_{e}} &= 1 - \sin^2 2 \theta_{ee} \sin^2 \left(\Delta m^{2}_{41} L/(4 E_{\nu}) \right)\,, \label{eq:tsurv31a} 
\end{align}

\noindent where $\sin^2 2 \theta_{\alpha \alpha} \equiv 4 \left( 1 - |U_{\alpha 4}|^{2} \right) |U_{
\alpha 4}|^{2}$, with $\alpha=e,\mu$ and $\sin^2 2\theta_{\mu e}\equiv 4|U_{e 4}|^{2}|U_{\mu 4}|^{2}$ are the oscillation amplitudes, defined by the elements of the $4\times 4$ generalized PMNS matrix elements $U_{e 4}$ and $U_{\mu 4}$, and $\Delta m^{2}_{41}$ is the squared mass difference between the fourth mass state $m_{4}$ (which is made majority by the sterile component of neutrino flavor basis) and the first mass state $m_{1}$. The probabilities in Eqs.~(\ref{eq:tconv31}), (\ref{eq:tsurv31}), (\ref{eq:tsurv31a}) at short-baselines depend on the three parameters $U_{e 4}, U_{\mu 4},\ \rm{and} \ \Delta m^{2}_{41}$~\cite{Peres:2000ic}.

\begin{figure}[t]
\centering
\includegraphics[scale=0.35]{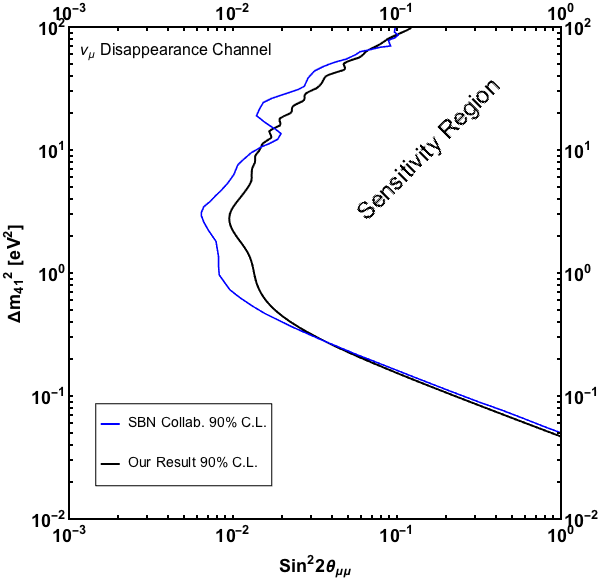}
\includegraphics[scale=0.35]{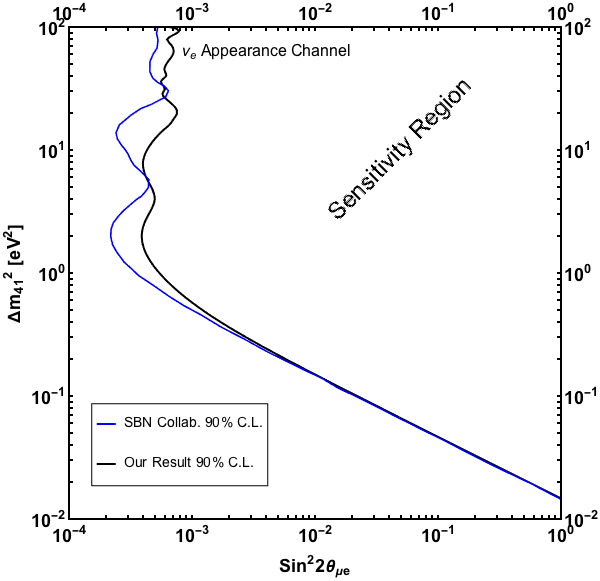}
\caption{Left (Right) panel: the sensitivity limit with 90\% C.L. for the 3+1 model for respectively the muon neutrino disappearance channel (electron neutrino appearance channel), in the parameter space  which depends on $\sin^2 2 \theta_{\mu \mu}$ ($\sin^2 2 \theta_{\mu e}$ )  and $\Delta m^{2}_{41}$. Exclusion (sensitivity) regions are to top-right of the black dashed curves in both panels. The solid black curve (solid blue curve) respectively shown our sensitivity (the SBN sensitivity was taken from Ref.~\cite{Antonello:2015lea}).
}
\label{fig:3+1-sens}
\end{figure}
\begin{figure}[h]
\centering
\includegraphics[scale=0.35]{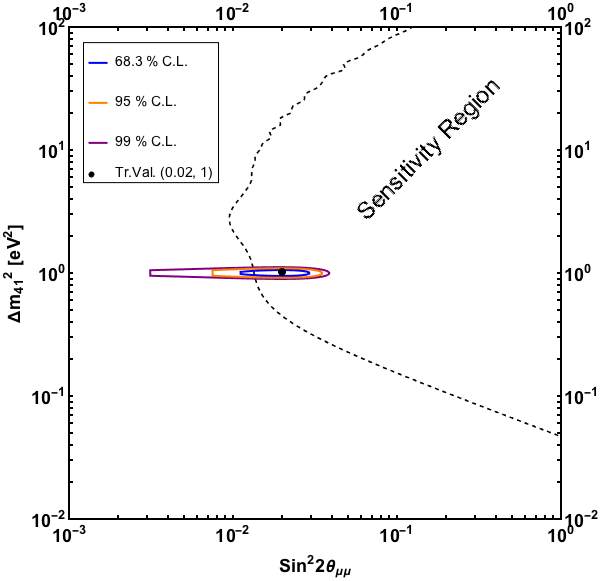}
\includegraphics[scale=0.35]{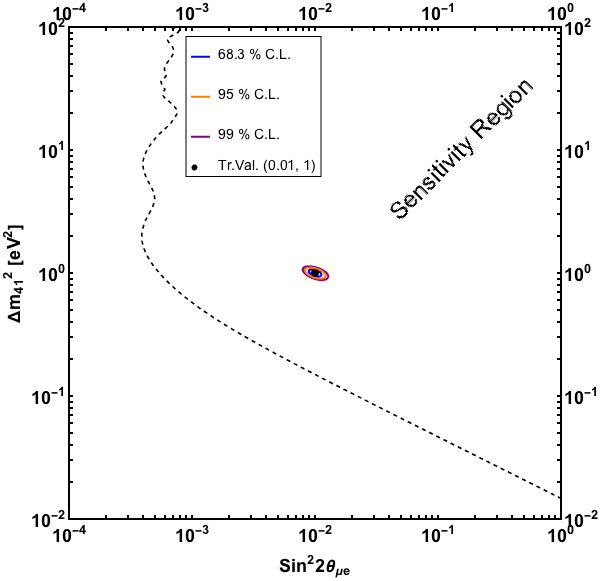}
\caption{Left (Right) panel: Allowed Regions considering the `true' neutrino event spectrum given by the $3+1$ model with the values $\sin^2 2 \theta_{\mu \mu} = 0.02$ and $\Delta m^{2}_{41} = 1$~eV$^2$ ( $\sin^2 2 \theta_{\mu e} = 0.01$ and $\Delta m^{2}_{41} = 1$~eV$^2$)  in the muon neutrino disappearance channel (the electron neutrino appearance channel). The dashed curve in both plots is the  sensitivity curve for the respective channels.}
\label{fig:3+1}
\end{figure}

We now test the two following cases in the 3+1 scenario:
\begin{enumerate}
\item  Assuming the `true' event energy distribution as compatible with the three-neutrino scenario and testing the 3+1 model. This gives the sensitivity of SBN to the 3+1 scenario that can be seen in Figure~\ref{fig:3+1-sens}. 
Exclusion regions are to the right of the black  curves for both appearance (right panel) and disappearance (left panel) channels.
We have a very good agreement with the SBN sensitivity, comparing the blue and solid curves in  Figure~\ref{fig:3+1-sens}.

\item Assuming as the `true' event energy distribution as compatible with the 3+1 scenario and testing the 3+1 model. This will give the accuracy of SBN facility to the parameters of the 3+1 scenario that can be seen in Figure~\ref{fig:3+1}.
For illustration purposes, we show the sensitivity as dashed black curves for the 3+1 model at the SBN from Figure~\ref{fig:3+1-sens}. The allowed regions assuming the `true' 3+1 parameters $\sin^2 2 \theta_{\mu \mu} = 0.02$, $\sin^2 2 \theta_{\mu e} = 0.01$ and $\Delta m^{2}_{41} = 1\,\text{eV}^2$ and also fitting 3+1 hypothesis. Notice that SBN is very sensitive to the square mass difference around $1\,\text{eV}^{2}$ and the precision that we can get for this value are very good and below 1\%. Even though not shown in the figure, large values of $\sin^2 2 \theta_{\mu \mu}$ and $\sin^2 2 \theta_{\mu e}$ gets more precise determined than the lower values shown in the plot. The fast oscillations  $\Delta m^{2}_{41} > 10$~eV$^{2}$ were handled assuming a  low-pass filter in our analysis using GLoBES 3.2.17 \cite{Huber:2004ka, Huber:2007ji}, otherwise we will have spurious results in our sensitivity for 3+1 model.
\end{enumerate}

\subsection{Discrimination power between LED scenario and the 3+1 scenario }
\label{discri}

\begin{figure}[t]
\centering
\includegraphics[scale=0.35]{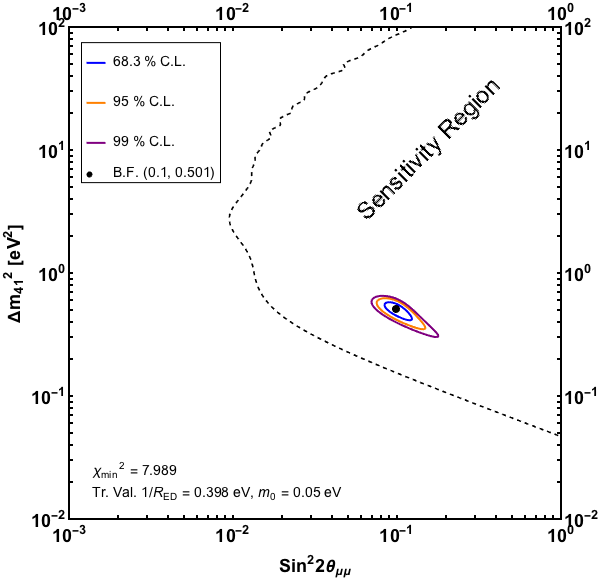}
\includegraphics[scale=0.35]{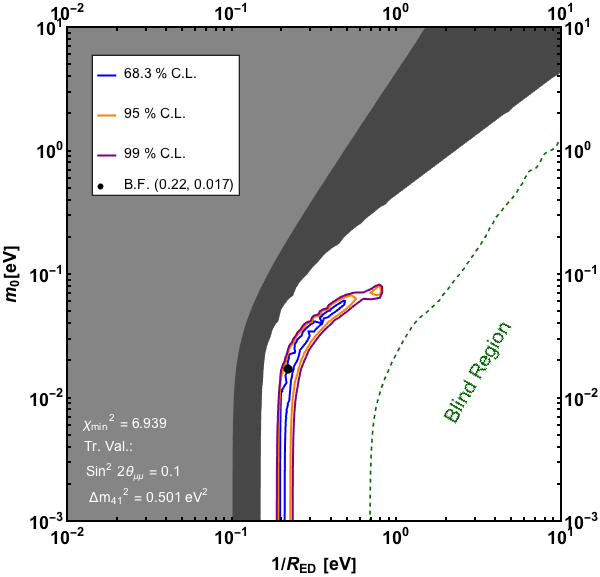}
\caption{Left panel, sensitivity results fitting the 3+1 model parameters assuming the `true' LED parameters $m_{0} = 0.05 \ \text{eV}$ and $1/R_{\rm{ED}} = 0.398\,\text{eV}$, for normal ordering. Right panel, sensitivity results fitting the LED parameters for the `true' 3+1 parameters $\sin^{2} 2\theta_{\mu \mu} = 0.1$ and $\Delta m_{41}^{2} =0.5\,\text{eV}^2$, also for normal ordering. The allowed sensitivity regions correspond to the $68.3\%$ of C.L. (blue), $95\%$ of C.L. (orange) and $99\%$ of C.L. (purple), the best-fit points appear as black dots.}
\label{fig:LED-3+1}
\end{figure}

One question that remains is, in the case SBN finds a departure from the three neutrino framework, is it possible to identify which of the two scenarios analyzed in this letter would be responsible for the new signal (assuming is not  something else)? In the following,  we analyze the discrimination power of the SBN experiment comparing both the LED and the 3+1 scenarios. 
Regarding the 3+1 fit to the LED scenario, we calculated events with the `true' LED parameters $m_{0} = 0.05 \ \text{eV}$ and $1/R_{\rm{ED}} = 0.398 \ \text{eV}$ assuming normal ordering. With this `true' events, both appearance and disappearance channels were fitted separately, fixing the parameters not shown in the plots. Figure~\ref{fig:LED-3+1} shows the result of the fit in the \emph{disappearance channel} (left panel) with allowed curves of $68.3\%$ of C.L. (blue), $95\%$ of C.L. (orange) and $99\%$ of C.L. (purple). The number of degrees of freedom (d.o.f.) was equal to $17$ (19 energy bins minus 2 free parameters). The best-fit of the test values is represented in the black dot and has values of $\sin^2 2 \theta_{\mu \mu} = 0.1$ and $\Delta m^{2}_{41} = 0.5 \ \text{eV}^2$. We have not found a good fit, where $\Delta \chi^2 = \chi^2_{3+1}-\chi^2_{\rm LED} = 8$ for the best-fit point, giving more than 2$\sigma$ of deviation between the two models.

We have also checked that when using the new set of parameters $m_{0} = 0.316 \ \text{eV}$ and $1/R_{\rm{ED}} = 1 \ \text{eV}$ for the muon disappearance case, we have obtained a $\Delta \chi^2 \approx 104$ for the best-fit (of the test values) point, implying a bad fit. This result can be explained due to the fact that for some values of the LED parameters, as in this case, more sterile states start to contribute in the oscillation probability and the 3+1 model cannot emulate the LED model.

Following a similar procedure, this time fitting the LED model for some `true' values for the 3+1 parameters, we could not obtain good fits. The analysis is shown in the right panel of Figure~\ref{fig:LED-3+1}. In fact, if we consider the amplitude $\sin^{2} 2\theta_{\mu \mu} = 0.01$ and the same $\Delta m_{41}^{2} = 0.5$~eV$^{2}$, the allowed regions would be almost entire inside the {\em Blind Region} (bottom-right part from the dashed green curve in the right panel of Figure~\ref{fig:LED-3+1}). From this analysis, we obtained the value $\Delta \chi^2 \approx 187$. 
We also considered the case of larger mixing with true values $\sin^{2} 2\theta_{\mu \mu} = 0.1$ and $\Delta m_{41}^{2} = 3$~eV$^{2}$ and we obtained the value $\Delta \chi^2 \approx 149$ for the best-fit point.

In the case of the \emph{electron neutrino appearance channel}, we repeated the same procedure done for the muon channel: we calculated events for a given `true' values for the LED parameters and we fitted the electron neutrino appearance parameters in the 3+1 model. The summary of the results are the following:

\begin{itemize}
\item For $m_{0} = 0.05$~eV and $1/R_{\rm{ED}} = 0.398$~eV, the best-fit and the allowed regions were located outside the Sensitivity Region with the value $\Delta \chi^2 \approx 78.3$ for the best-fit point, implying a very poor fit.
\item For $m_{0} = 0.316$~eV and $1/R_{\rm{ED}} = 1$~eV, the best-fit and allowed regions were located outside the sensitivity region, with $\Delta \chi^2 \approx 538$ for the best-fit point, implying a very poor fit.
\end{itemize}

\noindent The previous results (for the electron appearance case) were somehow expected since we could only obtain LED sensitivity from electron neutrino channel in Figure~\ref{fig:LED-fit} with effects of the LED parameters in the background. Then, we should not expect that the signal of the electron neutrino conversion can be fitted with the 3+1 parameters. In other words, evidence of electron appearance in short-baseline experiments would be inconsistent with LED hypothesis. Similar conclusion was made in Ref.~\cite{Carena:2017qhd}.

\begin{table}[t]
\centering
\smallskip
\begin{adjustbox}{width=1\textwidth}
\begin{tabular}{||lll||}
\toprule
 & $\nu_{\mu}$ Disappearance & $\nu_{e}$ Appearance \\
\midrule
\backslashbox{Test model}{True hypothesis} 
& LED $(m_0,1/R_{\rm ED})$ & LED $(m_0,1/R_{\rm ED})$ \\
\midrule
\multirow{3}{*}{3+1 $(\sin^2 2\theta_{\mu\mu} \ \rm{or} \ \sin^2 2\theta_{\mu e},\Delta m_{41}^2)$} & True: (0.05 eV, 0.398 eV) & True: (0.05 eV, 0.398 eV) \\
    & best fit test Values: (0.1, 0.5~eV$^2$) & - \\
    & $\Delta \chi^2 \approx$ 8 & $\Delta \chi^2 \approx$ 78 \\
\midrule
\multirow{3}{*}{3+1 $(\sin^2 2\theta_{\mu\mu} \ \rm{or} \ \sin^2 2\theta_{\mu e},\Delta m_{41}^2)$}& True: (0.316 eV, 1 eV) & True: (0.316 eV, 1 eV) \\
   & - & - \\  
   & $\Delta \chi^2 \approx$ 104 & $\Delta \chi^2 \approx$ 538 \\ 
  \hline \hline

\backslashbox{Test model}{True hypothesis} &  3+1 $(\sin^2 2\theta_{\mu\mu}, \Delta m_{41}^2)$ &  3+1 $(\sin^2 2\theta_{\mu e}, \Delta m_{41}^2)$ 
\\

\midrule
\multirow{3}{*}{LED $(m_0,1/R_{\rm ED})$} & True: (0.1, 0.5~eV$^2$) &  \\
    & best fir test values: (0.017 eV, 0.22 eV) & \ \ \ \ \ \ \ * \\
    & $\Delta \chi^2 \approx$ 6.8 &  \\

\midrule
  \multirow{3}{*}{LED $(m_0,1/R_{\rm ED})$} & True: (0.01, 0.5~eV$^2$) &  \\
    & - & \ \ \ \ \ \ \  * \\
    & $\Delta \chi^2 \approx$ 187 &  \\

\midrule
  \multirow{3}{*}{LED $(m_0,1/R_{\rm ED})$} & True: (0.1, 3~eV$^2$) &  \\
    & - & \ \ \ \ \ \ \ * \\
    & $\Delta \chi^2 \approx$ 149 &  \\
  \midrule
  \hline
\multicolumn{3}{l}{\footnotesize ( - ) Best-Fit Test Value is outside Exclusion Region, \ ( * ) LED does not expect positive signal of $\nu_{e}$ appearance in SBN.}
\end{tabular}
\end{adjustbox}
\caption{Discrimination power of SBN facility for 3+1 model and LED model.}
\label{summ}
\end{table}
The right panel of Figure~\ref{fig:LED-3+1} also shows the LED fit for a given set of `true' parameters of the $3+1$ model considering only muon disappearance. We fixed the 3+1 parameters $\sin^{2} 2\theta_{\mu \mu} = 0.1$ and $\Delta m_{41}^{2} =0.5$~eV$^{2}$ and fitted the LED parameters for normal ordering. The allowed curves corresponds to the $68.3\%$ of C.L. (blue), $95\%$ of C.L. (orange) and $99\%$ of C.L. (purple). The best-fit point obtained is $m_{0} = 0.017$~eV and $1/R_{\rm{ED}} = 0.22$~eV. Following the same procedure, we found $\Delta \chi^2 \approx 6.8$ for the best-fit point. 

As we discussed in Section~\ref{positive}, with information of the electron neutrino appearance channel (and not the muon disappearance) one can discriminate the LED scenario from the standard three-neutrino case only if changes in the background (i.e. the electron neutrino disappearance from the intrinsic $\nu_{e}$ of the beam) are considered. In this way, LED is not contributing to the {\em signal} ($\nu_{e}$ conversion) in the electron neutrino channel. Therefore, when regarding the LED fit under 3+1 scenario on these conditions, we would not expect to accommodate LED parameters for any set of `true' parameters of the $3+1$ model considering only the signal of electron neutrino appearance channel.

Finally, all the results obtained for the discrimination power of LED and the 3+1 model are summarized in Table~\ref{summ}.

\section{Summary and conclusions} \label{summary}
In the dawn of the new era of high precision neutrinos experiments, the search for Beyond Standard Model (BSM) physics will bring an understanding of the  mechanism beyond neutrino masses and neutrino mixing. The  possibility to have in Nature the presence of large extra dimension is intriguing and it has several consequences for the phenomenology of neutrino physics, such as the existence of infinite tower of Kaluza-Klein states of sterile neutrinos. The Short-Baseline Neutrino Program SBN at FERMILAB will fully test the presence of large extra dimension (LED) in neutrino oscillations. 

We have developed GLoBES simulation files~\cite{aedl-sbn} that include the three detectors at SBN facility where information of the two main channels of SBN program, the $\nu_{\mu}$ muon neutrino disappearance channel and the $\nu_{e}$ electron neutrino appearance channel, are included. In the paradigm of three neutrino oscillation, we expect to see no oscillation in any of SBN detectors.  
With the assumption that we measure no oscillations in any of SBN detectors, we can put bounds on the LED scenario. In the LED scenario, the non-standard oscillations are accounted for with two parameters, the lightest Dirac neutrino mass $m_0$ and the radius of large extra dimension $R_{\rm ED}$. We have shown in Figure~\ref{fig:prob} the regions with sizable muon neutrino disappearance probability and electron neutrino appearance probability in the presence of LED, for either normal or inverted hierarchy of active states. The typical values that we can test are $P(\nu_{\mu}\to \nu_{\mu})\sim 0.90 $ and $P(\nu_{\mu} \to \nu_{e})\sim 10^{-4}-10^{-3}$ for a $L/E_{\nu}=1.2$ km/GeV.

We showed in Figure~\ref{fig:LED-all} the sensitivity plot for the LED scenario that is the main result of this work, based on the simulation details described in Section~\ref{simulation}. The solid green curve is the sensitivity of LED scenario, the other dashed curves are the constraints/sensitivities from other experiments for LED scenario and the pink region is the allowed region to explain the reactor neutrino anomaly. We notice that SBN sensitivity curve has, for normal ordering, {\em the strongest bound} for almost all parameter region, with exception of the values of $m_0>2\times 10^{-1}$~eV and $1/R_{\rm ED}>3$~eV, for both orderings. From Figure~\ref{fig:LED-all}, we have learned that all sensitivity to LED scenario came from the  muon disappearance channel and that electron neutrino appearance channel plays a marginal role. 

Any positive signal of a neutrino oscillation in the SBN facility will be a departure of the present three neutrino paradigm. The main goal of the SBN facility is to test the hint of neutrino oscillation from LSND, Mini-Boone and reactor anomaly. This hint is more usually discussed in the context of the 3+1 scenario with one additional sterile neutrino. Then, we first reproduced the sensitivity region for both channels considered in this letter, under the 3+1 framework with the assumptions described in detail in Section~\ref{simulation}. Then, we computed the sensitivity region and compared it with the official sensitivity region of the SBN proposal, reaching a good agreement as shown in Figure~\ref{fig:3+1-sens}. In Figure~\ref{fig:3+1}, we showed the precision that we can have for a given choice of the parameters in a true 3+1 oscillation scenario. We found that the two channels provide sufficient information  to get a few  percent of accuracy in the oscillation parameters. 

Finally, the remaining question of the power discrimination of the SBN facility: Can the SBN be able to discriminate different physics scenarios when it has a clear departure from the three-neutrino paradigm in the data?, was answered. Table~\ref{summ} summarizes our results. It is possible to discriminate between both models at $3\sigma-10 \sigma$. The worst scenario was shown in Figure~\ref{fig:LED-3+1}, where we get a $2\sigma -3\sigma$ discrimination using the muon disappearance channel only. For other choices of parameters, as detailed in Table~\ref{summ}, we can easily discriminate the source of new physics in the SBN experiment, the large extra dimension or the 3+1 scenario.

\begin{acknowledgments}
 G.V.S is  thankful for the support of FAPESP funding Grant  No. 2016/00272-9 and No. 2017/12904-2. G.V.S. thanks the useful discussions with Pedro Pasquini and Andr\'e de Gouv\^ea. D. V. F.  is  thankful for the support of FAPESP funding Grant  No. 2017/01749-6. O.L.G.P. is thankful for the support of FAPESP funding Grant  No. 2014/19164-6,  No. 2016/08308-2, FAEPEX funding grant No. 519.292, CNPQ research fellowship No. 307269/2013-2 and No. 304715/2016-6. 
\end{acknowledgments}

\bibliographystyle{JHEP}
\bibliography{sbl-extra}


\end{document}